\documentclass{aa}
\usepackage{color,graphicx}
\usepackage{amsmath}
\usepackage{booktabs} 
\usepackage{natbib}
\usepackage[varg]{txfonts} 
\bibpunct[]{(}{)}{;}{a}{}{,} 

\begin{document}

\title{Time-distance helioseismology of solar Rossby waves\thanks{A movie is available in electronic form at
http://www.aanda.org}}
\titlerunning{Time-distance helioseismology of Rossby waves}

\author{
Zhi-Chao Liang \inst{\ref{mps}}
\and Laurent Gizon \inst{\ref{mps},\ref{gottingen},\ref{NYUAD}}
\and Aaron C. Birch \inst{\ref{mps}}
\and Thomas L. Duvall, Jr. \inst{\ref{mps}}
}

\institute{
Max-Planck-Institut f\"ur Sonnensystemforschung, Justus-von-Liebig-Weg 3, 37077 G\"ottingen, Germany\\
\email{zhichao@mps.mpg.de} \label{mps}
\and Institut f\"ur Astrophysik, Georg-August-Universit\"at G\"ottingen, Friedrich-Hund-Platz 1, 37077 G\"ottingen, Germany \label{gottingen}
\and Center for Space Science, NYUAD Institute, New York University Abu Dhabi, PO Box 129188, Abu Dhabi, UAE \label{NYUAD}
}

\date{Received $\langle$date$\rangle$ / Accepted $\langle$date$\rangle$}

\abstract
{ 
  Solar Rossby waves (r modes) have recently been discovered in the near-surface horizontal flow field  using the techniques of granulation-tracking and ring-diagram analysis applied to six years of SDO/HMI data.
} { 
  Here we apply time-distance helioseismology to the combined  SOHO/MDI and SDO/HMI data sets, which cover 21~years of  observations from May 1996 to April 2017.
  The goal of this study is to provide an independent confirmation over two solar cycles and in deeper layers of the Sun.
} { 
  We have measured south-north helioseismic travel times along the equator, which are sensitive to subsurface north-south flows.
  To reduce noise, the travel times were averaged over travel distances from 6$\degr$ to 30$\degr$; the mean distance corresponds to a p-mode lower turning point of $0.91$~$R_\odot$.
  The 21-year time series of travel-time measurements was split into three seven-year subsets and transformed to obtain power spectra in a corotating frame.
} { 
  The power spectra all show peaks near the frequencies of the classical sectoral Rossby waves for azimuthal wavenumbers in the range $3 \leq m \leq 15$.
  The mode frequencies and linewidths of the modes with $m \leq 9$ are consistent with a previous study whereas modes with $m \geq 10$ are shifted toward less negative frequencies by 10\,--\,20~nHz.
  While most of these modes have e-folding lifetimes on the order of a few months, the longest lived mode, $m=3$, has an e-folding lifetime of more than one year.
  For each mode, the rms velocity at the equator is in the range of 1\,--\,3~m~s$^{-1}$, with the largest values for $m\sim10$.
  No evidence for the $m = 2$ sectoral mode is found in the power spectrum, implying that the rms velocity of this mode is below $\sim$0.5~m~s$^{-1}$.
} { 
  This work confirms the existence of equatorial global Rossby waves in the solar interior over the past two solar cycles and shows that time-distance helioseismology is a promising technique to study them deep in the convection zone.
}


\maketitle
\section{Introduction}
Spheroidal oscillations associated with solar f, p, and g~modes have been studied intensively in the past few decades \citep[see, e.g., ][, for a recent review]{Basu2016}.
Toroidal oscillations, including r~modes, have been discussed in the literature as well \citep[e.g.,][]{Papaloizou1978, Unno1989}.
For these modes the Coriolis force is the dominant restoring force; they are similar to the Rossby waves observed in the Earth's atmosphere and oceans.
\citet{Provost1981}, \citet{Smeyers1981}, and \citet{Saio1982} derived linearized equations for the analysis of r~modes in uniformly and slowly rotating stars.
\citet{Wolff1986} applied the analysis of r~modes to the solar case and discussed the possibility of different radial orders.
Later, \citet{Wolff1998} studied the effect of differential rotation on r~modes and suggested that nonsectoral modes might be suppressed while sectoral modes are the least affected because their motions are concentrated in the equatorial regions.

In the corotating frame and to lowest order in the mean solar rotation rate $\Omega$, the frequencies of r~modes are given by
\begin{equation} \label{eq:ro}
  \omega \approx -\frac{2m\Omega}{\ell(\ell+1)} \quad \text{for} \quad m > 0,
\end{equation}
where $\ell$ is the harmonic degree, and $m$ is the azimuthal order.
A consequence of this is that the horizontal flow field arising from the r-mode oscillations characterized by $(\ell, m)$ drifts at a phase velocity of $\omega/m = -2\Omega/[\ell(\ell+1)] < 0$.
The negative phase velocity means the drift direction is retrograde.

\citet{Loeptien2018} used six years of observations from the Helioseismic and Magnetic Imager on board the Solar Dynamical Observatory \citep[SDO/HMI:][]{Scherrer2012,Schou2012} to provide a direct and unambiguous detection of Rossby waves at the surface and in the outer 20~Mm of the Sun.
Using both granulation-tracking \citep[e.g.,][]{Loeptien2016,Loeptien2017} and ring-diagram analysis \citep[e.g.,][]{Bogart2015}, \citet{Loeptien2018} find radial vorticity patterns along the equator which propagate retrograde in the corotating frame with a dispersion relation that is consistent with Eq.~\ref{eq:ro} for the $m=\ell$ case (sectoral modes).
\citet{Loeptien2018} find no evidence for the existence of the nonsectoral modes.

\begin{figure}
  \resizebox{\hsize}{!}{\includegraphics{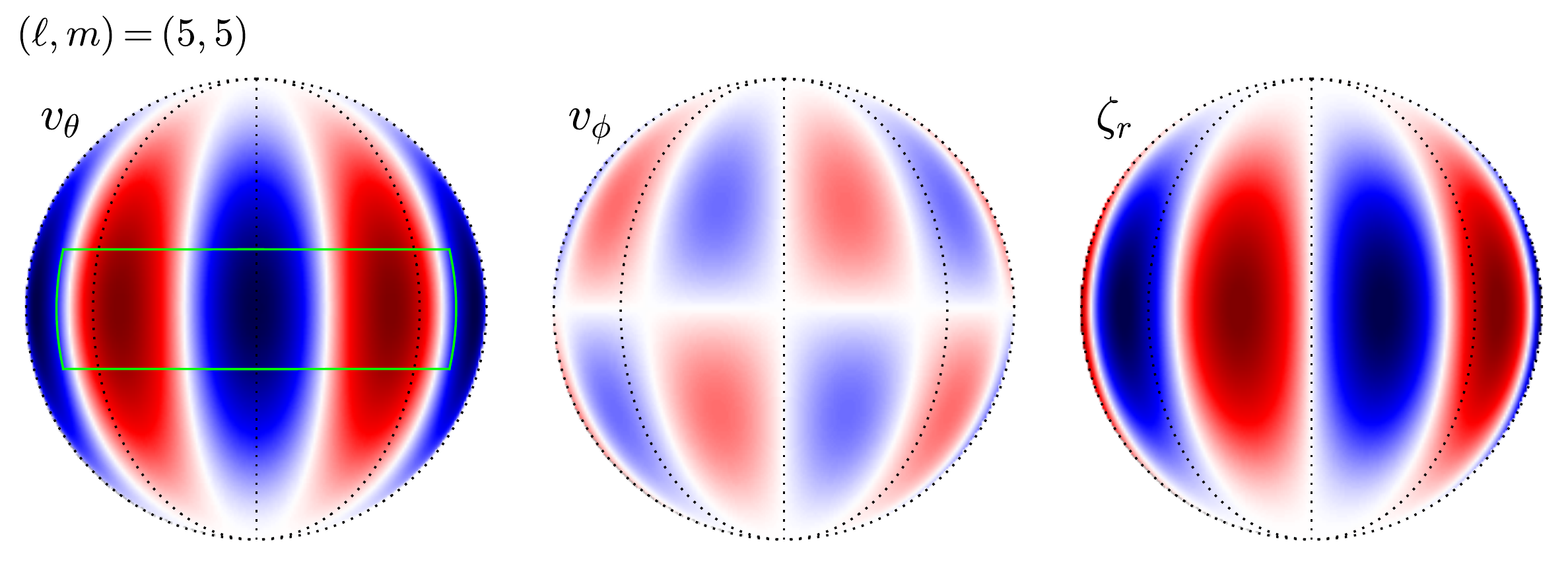}}
  \caption{ \label{fig:m5}
  Classical sectoral r mode with $m=5$, seen from the equatorial plane of a uniformly rotating solar model.
  The three panels show the southward flow $v_\theta$, the prograde flow $v_\phi$, and the radial vorticity $\zeta_r$ (from left to right) in the corotating frame.
  The color scale is the same for $v_\theta$ and $v_\phi$ with red positive and blue negative while the colors for the $\zeta_r$ indicate radially outward (red) or inward (blue) directions.
  The black dotted lines represent constant longitudes fixed in the corotating frame.
  The green rectangle marks the equatorial area ($\pm15\degr$) in which we measure $v_\theta$ in this paper.
  A movie showing the patterns propagating in the retrograde direction in the corotating frame is available online.
}
\end{figure}

The r-mode velocity for sectoral modes at the equator is solely in the north-south direction \citep[see][, for sketches of the flow field of r~modes]{Saio1982}.
Figure~\ref{fig:m5} shows the flow field and vorticity of classical r-mode oscillations for the case of $m=\ell=5$.
For this paper we have used time-distance helioseismology to measure subsurface flows in the meridional direction along the equator.
The time-distance helioseismology pipeline of \citet{Liang2018} applies not only to the recent SDO/HMI data but also to SOHO/MDI data \citep{Scherrer1995}.
The use of time-distance technique and the SOHO/MDI data would provide an independent confirmation of the findings of \citet{Loeptien2018}.

\section{Data analysis and results} \label{sec:data}

\subsection{Time-distance analysis} \label{sec:td}
We use medium-$\ell$ Dopplergrams taken by SOHO/MDI and SDO/HMI covering the period from May 1996 to April 2017.
Background Doppler signals, such as solar rotation, are removed by subtracting a one-hour running mean for each pixel.
A band-pass frequency filter is further applied to isolate the p~modes within 2\,--\,5~mHz.
Every 24-h time series of Dopplergrams are then projected into heliographic coordinates with a map scale of 0.6$\degr$~pixel$^{-1}$ and tracked at the Carrington rotation rate.
An improvement to the mapping procedure used in \citet{Liang2018} is made here by taking an error in the inclination angle of the solar rotation axis into account (see Appendix~\ref{app:dI}).

We compute the cross-covariance function (CCF) between pairs of points arranged in the arc-to-arc geometry used by \citet{Liang2018} where two 30$\degr$-wide concentric arcs, separated by an angular distance $\Delta$, are aligned in the north-south direction.
The CCFs between pairs of points on the opposing arcs are averaged and associated with the central point of the two arcs.
The procedure is repeated for different central points located within $\pm15\degr$~latitude at intervals of 0.6$\degr$ in longitude and latitude, and for the distance range $\Delta=6\degr$--$30\degr$ in steps of 0.6$\degr$.
Unlike the averaging scheme in \citet{Liang2018} where the CCFs were averaged over longitude and over days in each month, the daily CCFs are Gaussian smoothed with $\text{FWHM}=12\degr$ in longitude and latitude, and subsampled along the equator at intervals of 10$\degr$ in longitude.

\begin{figure}
  \centering
  \resizebox{0.7\hsize}{!}{\includegraphics{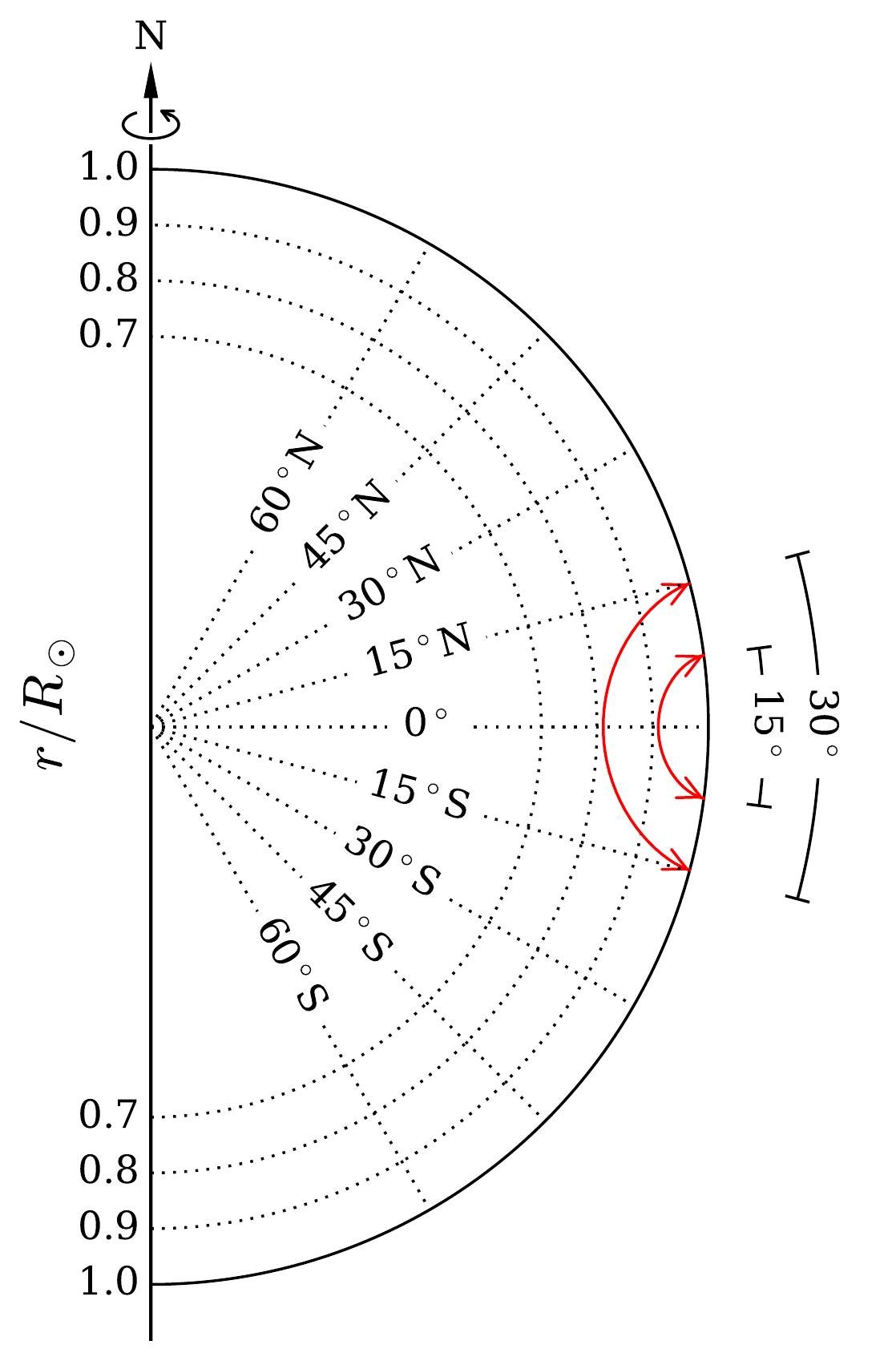}}
  \caption{ \label{fig:ray}
  Schematic plot of acoustic ray paths (red lines) that connect pairs of points across the equator in a meridional plane.
  The travel-time difference between the southward and northward propagating acoustic waves is sensitive to the north-south flow along the ray path.
  The larger the angular distance between the observation points at the surface, the deeper the lower turning point of the rays.
}
\end{figure}

The south-north travel-time shift in this work is defined as $\delta\tau = \tau_{\rm s} - \tau_{\rm n}$, where $\tau_{\rm n}$ and $\tau_{\rm s}$ are the northward and southward travel times of the first-skip wavelets in the CCF, respectively.
We measure the south-north travel-time shifts from the spatially smoothed CCFs using the linearized one-parameter fitting method \citep{Gizon2002} as this algorithm is more robust to noise \citep{Gizon2004}.
More precisely, a 20-min interval around the first-skip wavelet in the CCF for positive time lag (i.e., the wavelet traveling in the northward direction) is selected as the reference function to compute the weight function derived by \citet{Gizon2002}.
The weighted sum of the differences between the southward and northward wavelets in the CCF gives an estimate of the south-north travel-time shift.
The measured south-north travel-time shifts for different travel distances are sensitive to subsurface north-south flows around the equator as depicted in Fig.~\ref{fig:ray}.
Data points within active regions are included in the averaging of CCFs in order to reduce the noise level.
Also, the periods when the SOHO spacecraft was rotated by 180$\degr$, which were not used in \citet{Liang2018}, are used here for these alternate three-month gaps would result in strong leakage sidelobes in the Fourier domain.

To enhance the signal-to-noise ratio, we average the travel-time shifts over all travel distances with a weighting function that takes the noise correlation between different distances into account (see Appendix~\ref{app:wts}).
The weighted mean distance is about 14.6$\degr$, for which the corresponding depth of the lower turning point from the ray approximation is about 63~Mm.
The weighted average of the travel-time shifts suffers from a strong annual variation, mostly due to the center-to-limb effects \citep{Zhao2012}.\footnote{
Because of the inclination of the solar rotation axis to the ecliptic, the solar equator does not pass through disk center when the solar tilt angle ($B_0$ angle) is not zero.
Therefore the measurements along the equator are affected by center-to-limb effects, especially in spring and autumn.
}
To remove this time-varying background, we fit and subtract a periodic function from the measured travel-time shifts (see Appendix~\ref{app:bkg}).

\begin{figure}
  \resizebox{\hsize}{!}{\includegraphics{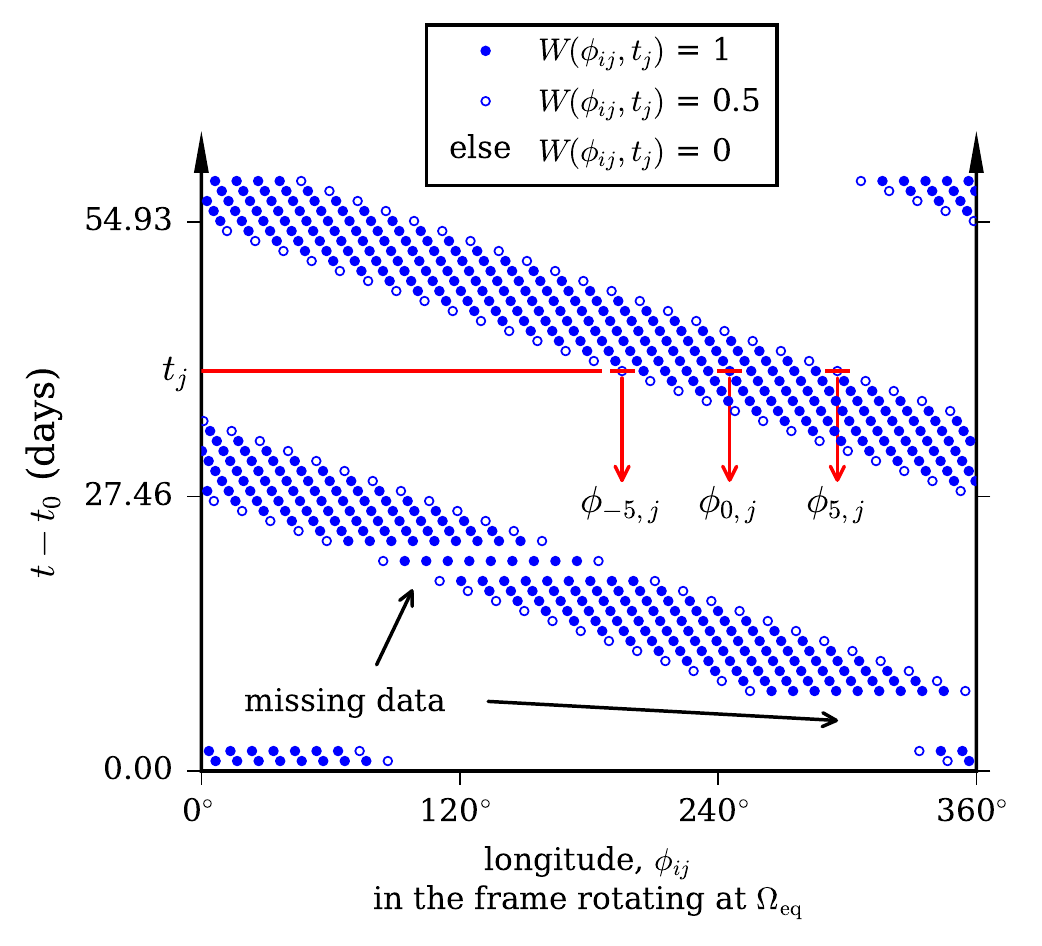}}
  \caption{ \label{fig:W}
  Section of the window function $W(\phi_{ij}, t_j)$  in the frame that rotates at $\Omega_{\rm eq}/2\pi=453.1$~nHz.
  At  time $t_j$ the longitudes of the travel times are denoted by $\phi_{ij}$ (see main text).
  The window function is equal to one for $i \in \{-4,\dots 4\}$, one half at the boundaries $i=\pm5$, and zero elsewhere.
  The window function is also zero for missing data.
  The temporal periodicity of the window is  $2\pi/(\Omega_{\rm eq} - \Omega_\oplus)=27.46$~days.
  }
\end{figure}

\begin{figure}
  \resizebox{\hsize}{!}{\includegraphics{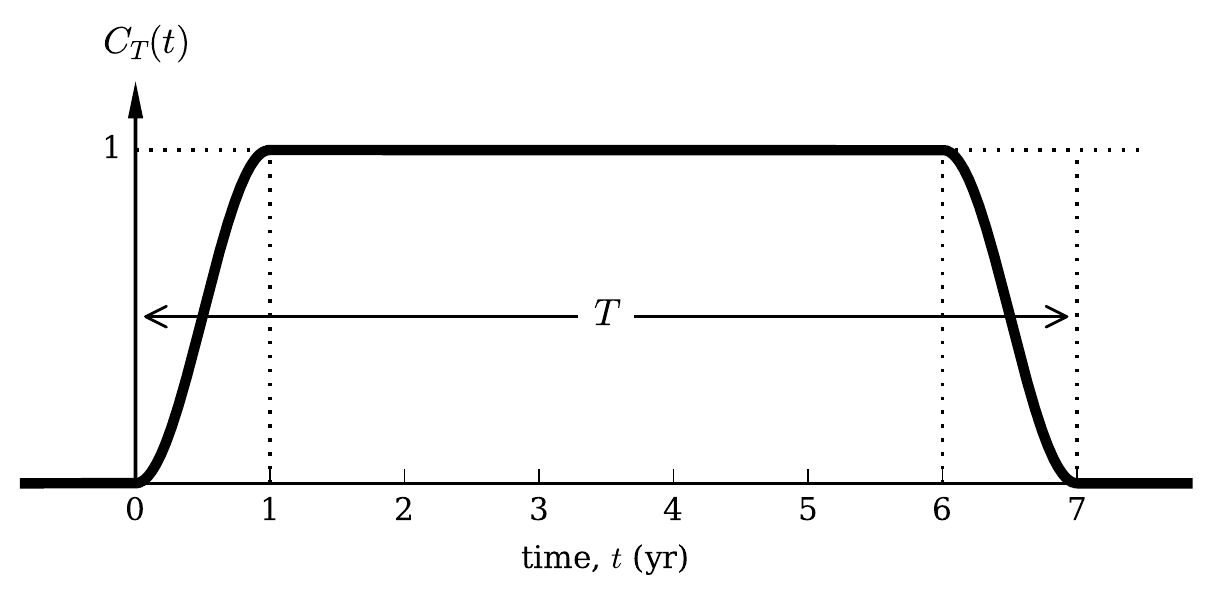}}
  \caption{ \label{fig:C}
 Window function $C_T(t)$ (thick solid line) that selects an observation period of $T=2556\Delta t \approx 7$~yr, tapered with one-year raised cosines at both ends.
  }
\end{figure}

The Carrington coordinate system rotates at $\Omega_{\rm cr}/2\pi=456.03$~nHz in a sidereal frame.
As viewed from the Earth, the Carrington coordinate system rotates at $\Omega_{\rm cr}-\Omega_\oplus$, where $\Omega_\oplus/2\pi=31.69$~nHz is the Earth's mean orbital frequency.
The central times of the daily measurements are denoted by  $t_j=t_0 + j\Delta t$, where $\Delta t=24$~h and $j \in \{0,1, \dots, 7669\}$, covering 21 years of data.
At each time $t_j$ the Carrington longitudes of the travel-time measurements are given by
\begin{equation} \label{eq:psi}
\psi_{ij} = \psi_{00} + i\Delta\psi - (\Omega_{\rm cr} - \Omega_\oplus)(t_j - t_0), 
\end{equation}
where $i \in \{-5,-4,\dots, 5\}$ (11 measurements along the equator each day) and  $\Delta\psi = 10^\circ$ is the spatial sampling rate. The longitude $\psi_{00}$ is the Carrington longitude of the central meridian as seen by the observer at time $t_0$, which is  the time when we start using MDI observations:
\begin{equation}
t_0 = \text{1 May 1996 12:00:00\_TAI} .
\end{equation}

To compare with the results of \citet{Loeptien2018}, we transform to a frame  that rotates at $\Omega_{\rm eq}/2\pi=453.1$~nHz, that is the solar surface equatorial rotation rate in the sidereal frame measured by global-mode helioseismology. The transformation between the two coordinate systems is given by
\begin{equation} \label{eq:trans}
\phi_{ij} = \psi_{ij} + (\Omega_{\rm cr} - \Omega_{\rm eq})(t_j-t_0),
\end{equation}
where $\phi_{ij}$ is the longitude measured in the new frame.
Figure~\ref{fig:W} is a  schematic plot of the corresponding window function $W(\phi_{ij}, t_j)$, which is equal to 1 for $i\in \{-4, \dots, 4\}$, tapered to $1/2$ at the boundaries $i=\pm 5$ where noise is slightly higher, and zero elsewhere. The window function is also zero for missing data. 

We denote by $\delta\tau(\phi_{ij},t_j)$ the travel times interpolated at the longitudes $\phi_{ij}$ in the new frame. In practice, we implement the interpolation in the spatial Fourier domain by multiplying the spatial Fourier transform of the travel times by the phase factor $\exp[-\mathrm{i}m(\Omega_{\rm cr}-\Omega_{\rm eq})(t_j-t_0)]$, where $m$ is the azimuthal wavenumber.

\begin{figure}[t]
  \resizebox{\hsize}{!}{\includegraphics{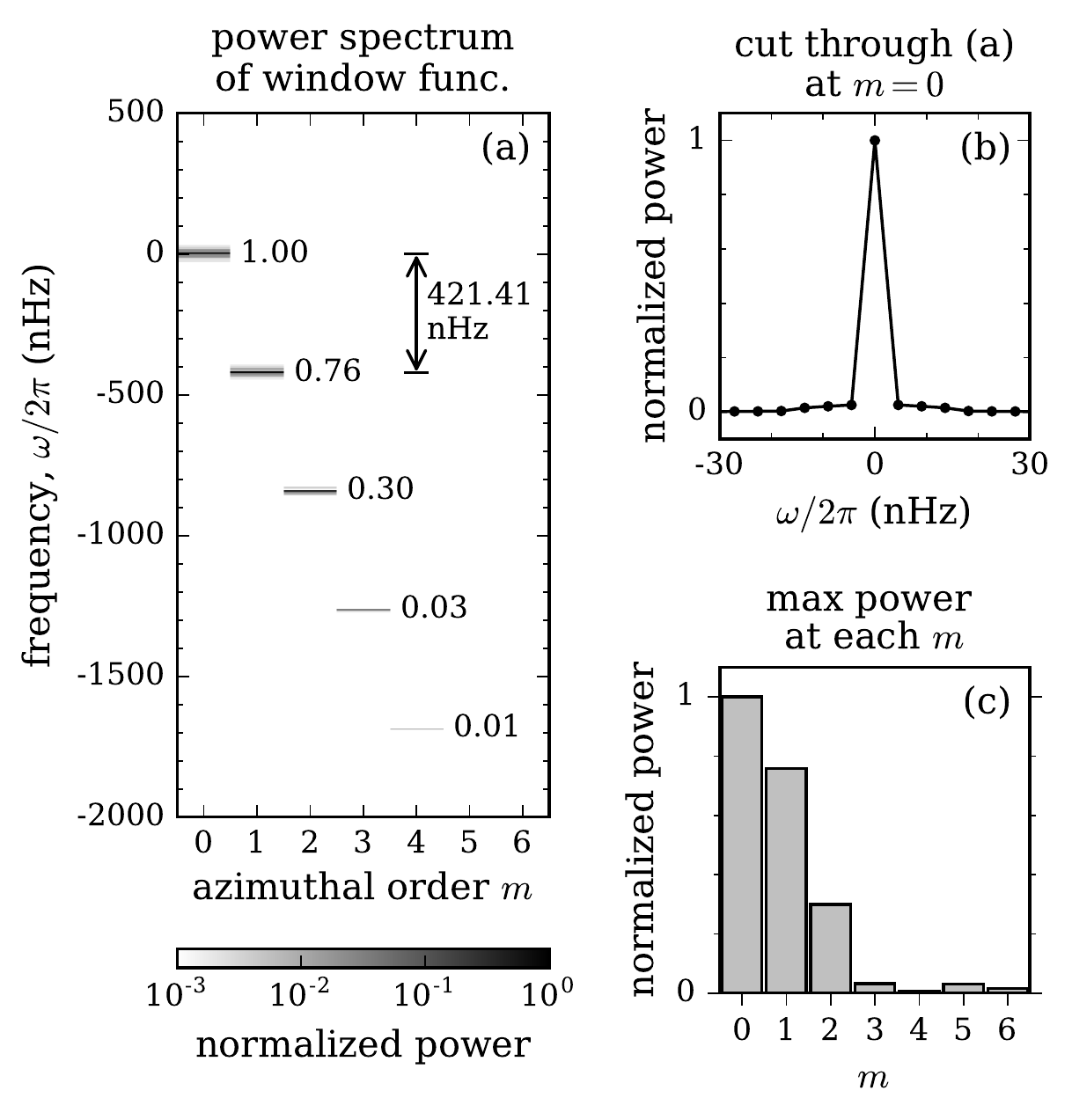}}
  \caption{ \label{fig:leak}
  Panel (a) shows the mean power spectrum of the window functions $W_T^{(k)}$.
  The power spectrum is normalized to unity at $(m,\omega)=(0,0)$.
  The maximum value of each peak is written right next to the peak.
  The frequency offset between the peaks with adjacent $m$ is $(\Omega_{\rm eq} - \Omega_\oplus)/2\pi=421.41$~nHz.
  Panel (b) shows a cut at $m=0$ through the power spectrum from panel (a).
  Panel (c) shows the maximum power of the peaks as a function of $m$, depicting the spatial leakage.
  }
\end{figure}

\subsection{Power spectra}
The full data set $\delta\tau(\phi_{ij},t_j)$ is divided into three consecutive periods of seven years.
The starting times of these three periods, denoted by $t^{(k)}$ with $k=\{1,2,3\}$, are given by
\begin{align*}
  t^{(1)} &=  t_0, \\
  t^{(2)} &= \text{1 May 2003 12:00:00\_TAI}, \\
  t^{(3)} &= \text{1 May 2010 12:00:00\_TAI} .
\end{align*}
The window functions for each  7-yr period is
\begin{equation}
    W_T^{(k)}(\phi_{ij}, t_j) = C_T(t_j-t^{(k)}) W(\phi_{ij}, t_j),
    \quad k=\{1,2,3\},
\end{equation}
where $C_T(t_j-t^{(k)})$ is a taper time window that selects an observation period of $T=N\Delta t\approx 7$~yr where $N=2556$, starting from time $t^{(k)}$ (see Fig.~\ref{fig:C}).

\begin{figure*}[ht]
  \centering
  \includegraphics[width=17cm]{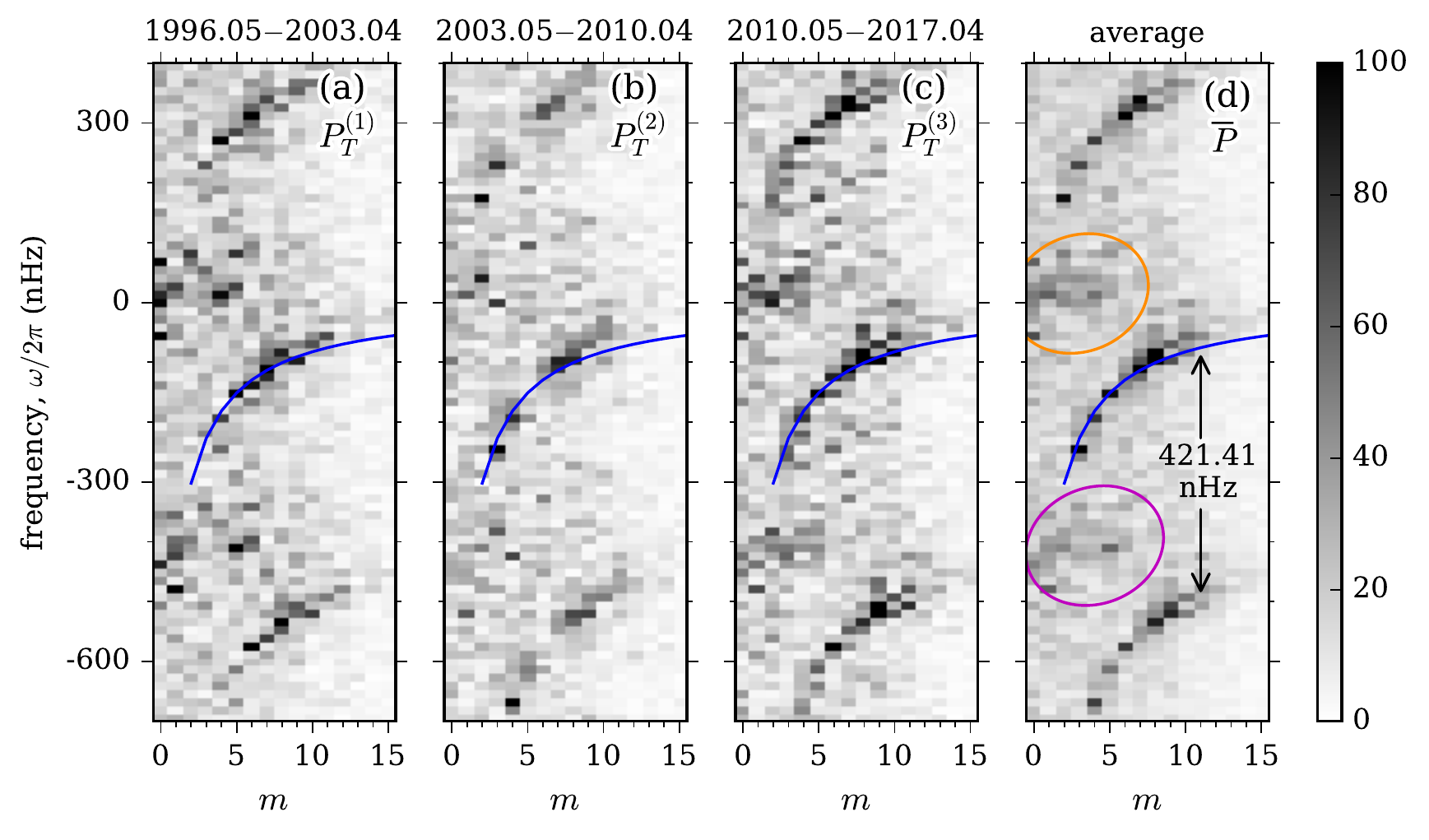}
  \caption{ \label{fig:pows}
  Power spectra of south-north travel-time shifts measured in the frame rotating at $\Omega_{\rm eq}/2\pi=453.1$~nHz.
  Panels (a) to (c) show $P_T^{(k)}(m, \omega)$ from three time periods.
  Panel (d) shows the mean spectrum $\overline P(m, \omega)$ from an average of the three $P_T^{(k)}(m, \omega)$ (seven years each).
  The blue lines highlight the dispersion relation of the classical Rossby waves described in Eq.~\ref{eq:ro} with $\ell = m$ and $\Omega = \Omega_\mathrm{eq}$.
  The orange ellipse marks the excess low-frequency power at low $m$ and the purple ellipse marks the spectral leakage from the low-frequency power.
  The gray scale is the same for the four panels and is shown in the color bar on the right.
  For clarity, the spectra are rebinned in frequency by a factor of three, such that the frequency resolution is $3/T=13.6$~nHz.
  }
\end{figure*}

The power spectrum of each 7-yr period is  
\begin{equation}
    P_T^{(k)}(m,\omega) = \frac{1}{N}\left|\sum_{i,j}
    W_T^{(k)}(\phi_{ij}, t_j) \;
    \delta\tau(\phi_{ij}, t_j) \;
    e^{-\mathrm{i}m\phi_{ij}+\mathrm{i}\omega t_j}\right|^2,
\end{equation}
where the azimuthal order $m$ is in the range $|m|\leq 18$ (spatial Nyquist frequency). For each $P_T^{(k)}(m,\omega)$, the temporal Nyquist frequency is 5787~nHz and the frequency resolution is $4.5$~nHz.

We denote the 2D Fourier transform of $W_T^{(k)}(\phi_{ij}, t_j)$ by $\widehat{W}_T^{(k)}(m,\omega)$.
Because the resulting $P_T^{(k)}(m,\omega)$ involves the convolution of $\widehat{W}_T^{(k)}(m,\omega)$, it is necessary to investigate the power spectrum of the window function, $|\widehat{W}_T^{(k)}(m,\omega)|^2$, from which the spectral leakage arises.
We note that $\widehat{W}_T^{(k)}(-m,-\omega) = \widehat{W}_T^{(k)*}(m,\omega)$ since the window function is real.

The mean power spectrum of the window functions, $\sum_{k=1}^3|\widehat{W}_T^{(k)}(m,\omega)|^2/3$, consists of well-defined peaks, as can be seen in Fig.~\ref{fig:leak}a  (plot restricted to $m \geq 0$).
The maximum values of the peaks at $m= 1$ and $m=2$ are 76\% and 30\% of the peak at the origin $(m,\omega)=(0,0)$.
The maximum power of the other peaks drops rapidly, 3\% for $m=3$ and 1\% for $m=4$.
The decrease in the peak power with increasing $m$ (as shown in Fig.~\ref{fig:leak}c) corresponds to that of a square window covering 100$\degr$ longitudes.
In Fig.~\ref{fig:leak}a, adjacent peaks are shifted in frequency by $(\Omega_{\rm eq}-\Omega_\oplus)/2\pi=421.41$~nHz due to the choice of coordinate frame.
Since the power spectrum $P_T^{(k)}$ involves a convolution in Fourier space of the solar data with $\widehat{W}_T^{(k)}$, the power spectrum of the data at $(m,\omega)$ leaks to multiple sidelobes at $\big(m+\delta m, \omega-\delta m\,(\Omega_{\rm eq}-\Omega_\oplus)\big)$, where $|\delta m| = 1, 2,\dots$.

Figure~\ref{fig:leak}b shows a cut at $m=0$ through the power spectrum of the window function.
The width of the main lobe along the $\omega$-axis is just one frequency bin and the frequency leaks to the neighboring frequency bins are nearly zero (two order of magnitude smaller than the central peak).
Since we observe only a fraction of the Sun (see Fig.~\ref{fig:W}), there is spatial leakage to the neighboring $m$ at frequency separations that are integer multiples of $421.41$~nHz.
Figure~\ref{fig:leak}c shows these leaks, which implies a resolution in $m$ of about $4$.

Figure~\ref{fig:pows} shows the $P_T^{(k)}(m, \omega)$ for the three different periods, as well as the mean spectrum
\begin{equation}
\overline{P}(m, \omega) = \frac{1}{3}\left(P_T^{(1)} (m, \omega)+P_T^{(2)} (m, \omega)+P_T^{(3)} (m, \omega)\right).
\end{equation}
The power distribution in all three $P_T^{(k)}(m, \omega)$ peaks around the eigenfrequencies of r~modes as described in Eq.~\ref{eq:ro} for the case of $\ell = m \geq 3$.
However, there is no clear sign of $\ell = m \leq 2$.
It is interesting to note that the mode amplitudes vary among the three $P_T^{(k)}(m, \omega)$, suggesting a possible temporal evolution of Rossby waves over the solar cycles.
After the averaging, the ridge structure in the $\overline{P}(m, \omega)$ along $\omega = -2\Omega_\mathrm{eq}/(m+1)$ becomes more prominent.
The other two ridges of power seen in Fig.~\ref{fig:pows}, which are separated from the central ridge by $(\delta m, \delta\omega) = \big(\pm 1, \mp (\Omega_{\rm eq}-\Omega_\oplus) \big)$ with a reduced amplitude ($60\sim80$\%), are due to the spectral leakage discussed previously.
Excess power at low frequencies is present at low $m$, which also leaks and modulates the spectrum at frequencies between $-300$ and $-500$~nHz.
It might be caused by large-scale convection or local flows around active regions that corotate with the Sun, and should not be mistaken for oscillation power.
While the annual variation ($31.7$~nHz) in the background has been removed in the analysis, harmonics at $\pm63.4$~nHz remain for $m=0$ as the background fitting in Appendix~\ref{app:bkg} only accounts for the first order term of $B_0$-angle variation.
We do not expect any aliasing at high $m$ since we applied a Gaussian smoothing with ${\rm FWHM=12^\circ}$ in longitude as mentioned in Sect.~\ref{sec:td}.

\begin{figure}[t]
  \resizebox{\hsize}{!}{\includegraphics{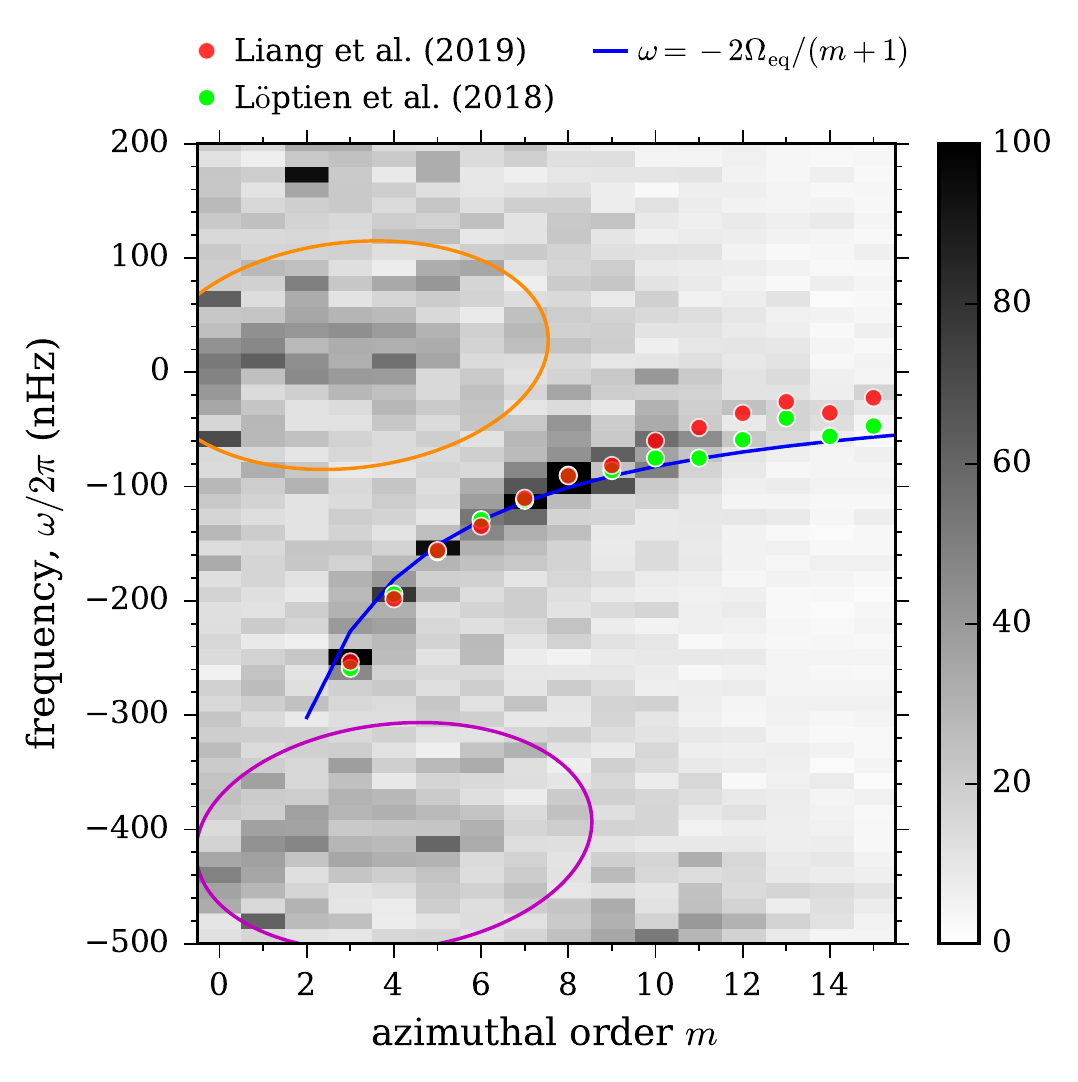}}
  \caption{ \label{fig:zoom}
  Enlargement of Fig.~\ref{fig:pows}d around the frequency range of interest (see Fig.~\ref{fig:fit} for the line profiles of individual modes in the range $3 \leq m \leq 14$).
  The red circles show the mode frequencies estimated from Lorentzian fits (see Sect.~\ref{sec:fit}).
  The errors in the mode frequencies are given roughly by the size of the red circles (see errors in Table~\ref{tab:fit}).
  For comparison, the mode frequencies measured by \citet{Loeptien2018} are also indicated by green circles.
  }
\end{figure}

Figure~\ref{fig:zoom} shows the $\overline{P}(m,\omega)$ in close-up around the frequency range of interest.
The distribution of r-mode power seems to shift toward less negative frequencies by 10--20~nHz for $m \geq 10$ compared to the mode frequencies measured by \citet{Loeptien2018}.

We note that our analysis is independent of the choice of reference frame.
To illustrate this point, we present the power spectrum of data computed in the frame of the observer (as seen from Earth) in Appendix~\ref{app:earth}.
Compared to the power spectrum computed in the corotating frame, this power spectrum is shifted by $m\times 421.41$~nHz for each $m$ value.
Figure~\ref{fig:mw1} displays a set of horizontal segments of size set by the resolution in $m$ ($\approx 4$, see Fig.~\ref{fig:leak}c).
While the resolution in $m$ is not one, one can easily retrieve the mode frequencies since the sectoral Rossby modes are well separated in frequency due to their long lifetimes.

\subsection{Mode frequencies and linewidths of r~modes} \label{sec:fit}
\begin{figure*}[h!]
  \centering
  \includegraphics[width=17cm]{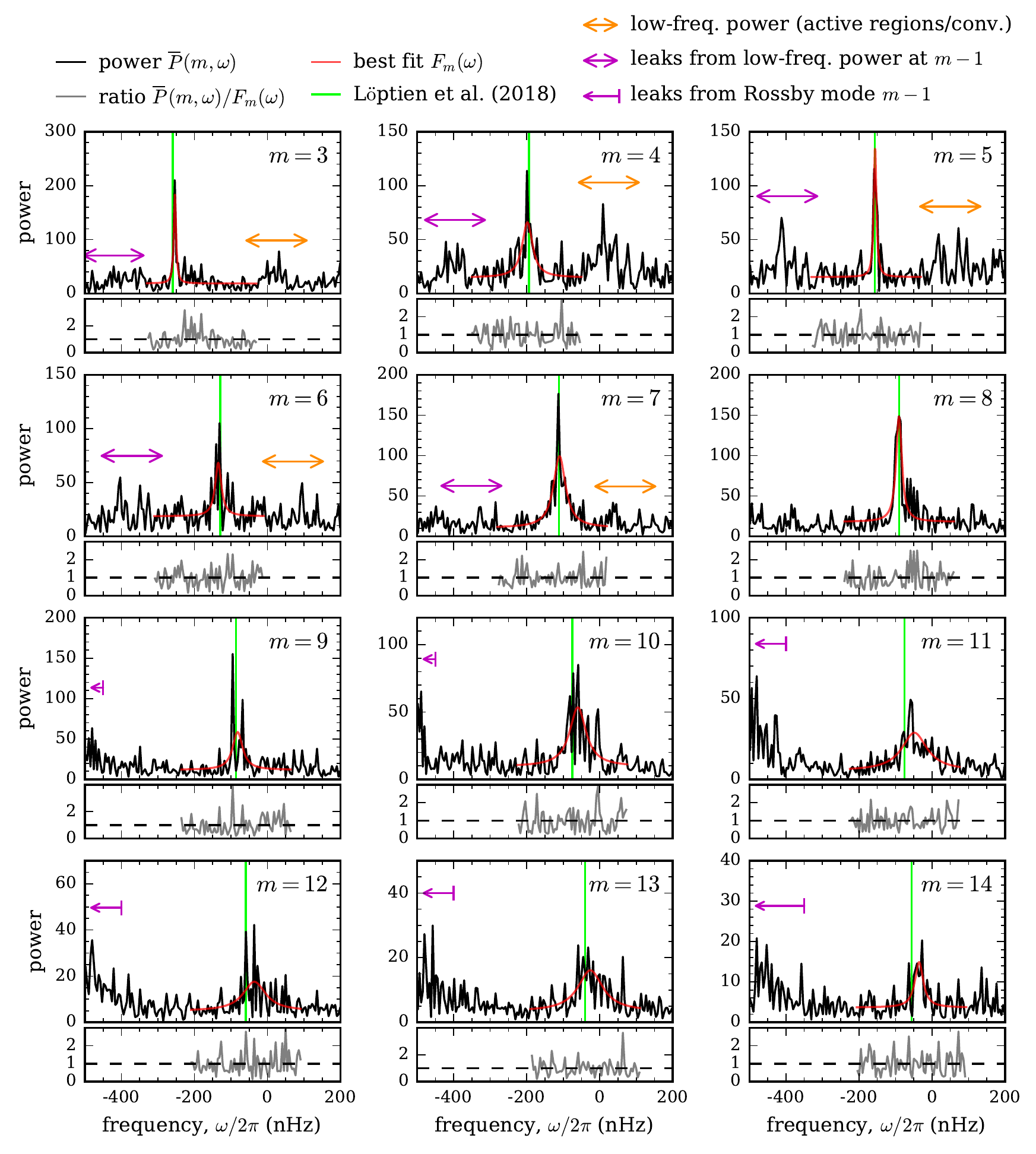}
  \caption{ \label{fig:fit}
  Power spectra of south-north travel-time shifts $\overline P(m, \omega)$ for modes in the range $3 \leq m \leq 14$ (black curves, with the frequency resolution $1/T=4.5$~nHz).
  The travel times are measured in the frame rotating at equatorial rotation rate $\Omega_\mathrm{eq}/2\pi=453.1$~nHz.
  The red lines are the fits $F_m(\omega)$ given by Eq.~(\ref{eq:fit}).
  The green vertical lines indicate the mode frequencies from \citet{Loeptien2018}.
  The orange arrows mark the excess low-frequency power at low $m$ that might be caused by active regions or large-scale convection.
  The purple arrows mark the leaks from the $m-1$ r-mode power and the low-frequency power at $m-1$.
  Each power spectrum is accompanied by a plot of the ratio $\overline P(m, \omega)/F_m(\omega)$ in gray in the lower panel, which is expected to have a mean of unity (dashed line) and a constant variance if the fit is not biased.
  }
\end{figure*}

To quantify the line profiles of r~modes, a model consisting of a Lorentzian function plus a constant background,
\begin{equation} \label{eq:fit}
  F_m(\omega) = \frac{A}{1 + [(\omega - \omega_m)/(\Gamma_m/2)]^2} + B,
\end{equation}
is fit to the $\overline P(m, \omega)$ at individual $m$.
Here $A$ is the maximum height of the Lorentzian function, $\omega_m$ the mode frequency, $\Gamma_m$ the full width at half maximum, and $B$ the constant background power.
To ensure that the Lorentzian fits are not affected by the spatial leaks from neighboring $m$ (see Fig.~\ref{fig:fit}, at negative frequencies, on the left) nor by the low-frequency power from active regions or convection (near zero frequency, on the right), we chose a fitting range [$\nu_{\rm start}$, $\nu_{\rm start}$+300~nHz], with $\nu_{\rm start}$ given in Table~\ref{tab:fit}.

The best fit is obtained by minimizing the sum
\begin{equation} \label{eq:mle}
  \sum_\omega \left( \ln F_m(\omega) + \frac{\overline{P}(m,\omega)}{F_m(\omega)} \right)
\end{equation}
with respect to the model parameters $A, B, \omega_m, \Gamma_m$ \citep{Duvall1986}.
Although Eq.~\ref{eq:mle} was derived assuming that the quantity $2\overline P(m, \omega)/F_m(\omega)$ has a chi-squared distribution with two degrees of freedom for a single realization, we can minimize the same function for the case of multiple realizations \citep{Anderson1990} as the $\overline{P}(m, \omega)$ is an average over three power spectra.
The minimization algorithm used is the downhill simplex method \citep[e.g.,][, Sect.~10.4]{Press1992}.
Table~\ref{tab:fit} lists the resulting best fit parameters for $3 \leq m \leq 15$.

Figure~\ref{fig:fit} shows the fitted $F_m(\omega)$ on top of the $\overline{P}(m, \omega)$, along with the results of \citet{Loeptien2018} for comparison.
We have also computed the spectrum without splitting the 21-yr-long time series, the results of which appear to be similar to Fig.~\ref{fig:fit} but with better frequency resolution and higher noise level.
The fitted mode frequencies and linewidths are remarkably consistent with that of \citet{Loeptien2018} for $3 \leq m \leq 9$.
However, excess power on the right of the main peak is present for $m \geq 8$, the extent of which for $m \geq 10$ becomes so large that the overall profiles shift toward less negative frequencies.

\begin{table*}[ht]
  \caption{Measured characteristics of solar sectoral r~modes.} 
  \centering

\begin{tabular}{c c c c c c c c c c c}
  \hline\hline
      & \multicolumn{9}{c}{This work}               & \citet{Loeptien2018} \\
      & \multicolumn{9}{c}{1996--2017 (MDI \& HMI)} & 2010--2016 (HMI) \\
  \cmidrule(lr){2-10} \cmidrule(lr){11-11}
  \vspace{-2pt}
      &                   &                 &                 &     &     & period in               & e-folding &                        &               &                 \\
  $m$ & $\nu_{\rm start}$ & $\omega_m/2\pi$ & $\Gamma_m/2\pi$ & $A$ & $B$ & $\Omega_{\rm eq}$ frame & lifetime  & $\delta\tau_{\rm rms}$ & $v_{\rm rms}$ & $\omega_m/2\pi$ \\
      & (nHz)             & (nHz)           & (nHz)           &     &     & (days)                  & (days)    & (s)                    & (m~s$^{-1}$)  & (nHz)           \\
  \hline
 3\rule[-1ex]{0pt}{3.2ex} & $-330$ & $-253 \pm 2$     & $7_{-3}^{+4}$    & $165_{-77}^{+173}$ & $18 \pm 1$     & $ -46$ & $505$ & $0.07$ & $1.0$ & $-259$           \\
 4\rule[-1ex]{0pt}{3.2ex} & $-350$ & $-198 \pm 5$     & $34_{-13}^{+15}$ & $51_{-14}^{+27}$   & $15_{-2}^{+1}$ & $ -58$ & $109$ & $0.08$ & $1.3$ & $-194_{-4}^{+5}$ \\
 5\rule[-1ex]{0pt}{3.2ex} & $-330$ & $-156 \pm 2$     & $8_{-4}^{+5}$    & $123_{-56}^{+111}$ & $15 \pm 1$     & $ -74$ & $445$ & $0.06$ & $1.0$ & $-157\pm4$       \\
 6\rule[-1ex]{0pt}{3.2ex} & $-310$ & $-135_{-5}^{+4}$ & $20_{-11}^{+12}$ & $50_{-17}^{+48}$   & $19_{-2}^{+1}$ & $ -86$ & $181$ & $0.06$ & $1.1$ & $-129\pm8$       \\
 7\rule[-1ex]{0pt}{3.2ex} & $-280$ & $-110 \pm 4$     & $40_{-10}^{+12}$ & $89_{-22}^{+31}$   & $11 \pm 1$     & $-105$ & $ 92$ & $0.11$ & $2.2$ & $-112\pm4$       \\
 8\rule[-1ex]{0pt}{3.2ex} & $-240$ & $-91 \pm 3$      & $19_{-6}^{+7}$   & $131_{-44}^{+72}$  & $18_{-2}^{+1}$ & $-128$ & $198$ & $0.09$ & $1.9$ & $-90\pm3$        \\
 9\rule[-1ex]{0pt}{3.2ex} & $-236$ & $-82 \pm 5$      & $33_{-12}^{+13}$ & $47_{-13}^{+24}$   & $12 \pm 1$     & $-142$ & $112$ & $0.08$ & $1.7$ & $-86\pm6$        \\
10\rule[-1ex]{0pt}{3.2ex} & $-225$ & $-60_{-6}^{+5}$  & $54_{-16}^{+18}$ & $44_{-10}^{+16}$   & $9 \pm 1$      & $-193$ & $ 68$ & $0.09$ & $2.2$ & $-75\pm5$        \\
11\rule[-1ex]{0pt}{3.2ex} & $-225$ & $-48 \pm 7$      & $84_{-24}^{+27}$ & $24_{-4}^{+7}$     & $5 \pm 1$      & $-239$ & $ 44$ & $0.08$ & $2.2$ & $-75\pm7$        \\
12\rule[-1ex]{0pt}{3.2ex} & $-209$ & $-36 \pm 8$      & $75_{-28}^{+31}$ & $13_{-2}^{+5}$     & $5 \pm 1$      & $-323$ & $ 49$ & $0.06$ & $1.7$ & $-59\pm6$        \\
13\rule[-1ex]{0pt}{3.2ex} & $-190$ & $-26 \pm 7$      & $82_{-25}^{+29}$ & $13_{-2}^{+4}$     & $3 \pm 1$      & $-449$ & $ 45$ & $0.06$ & $2.0$ & $-40\pm10$       \\
14\rule[-1ex]{0pt}{3.2ex} & $-206$ & $-35 \pm 5$      & $27_{-12}^{+13}$ & $11_{-4}^{+8}$     & $4 \pm 0.4$    & $-326$ & $135$ & $0.03$ & $1.2$ & $-56_{-7}^{+6}$  \\
15\rule[-1ex]{0pt}{3.2ex} & $-197$ & $-22_{-3}^{+2}$  & $11_{-5}^{+7}$   & $19_{-8}^{+19}$    & $4 \pm 0.3$    & $-519$ & $345$ & $0.03$ & $1.1$ & $-47_{-6}^{+7}$  \\
  \hline
\end{tabular}

\tablefoot{ \label{tab:fit} 
  Parameters $\omega_m$ (mode frequency), $\Gamma_m$ (full width at half maximum), $A$ (maximum height of the Lorentzian function), and $B$ (background power) that give the best fit to the power spectrum $\overline{P}(m,\omega)$ for each $m$ in the range $3 \leq m \leq 15$.
  The frequency $\nu_{\rm start}$ defines the frequency range [$\nu_{\rm start}$, $\nu_{\rm start}$+300~nHz] for the Lorentzian fit.
  The 68\% confidence interval on each parameter is indicated by the upper and lower bounds estimated from Monte Carlo simulations.
  The modes with $m < 10$ have a quality factor $ \omega_{m}/\Gamma_m > 1$ in the corotating frame (the frame for Rossby waves).  However, the modes with $m \geq 10$ have a quality factor less than or comparable to one.
  The e-folding lifetime, $2/\Gamma_m$, is given for each mode, as well as the oscillation period measured in the Sun's corotating frame (rotating at $\Omega_{\rm eq}/2\pi=453.1$~nHz).
  Negative oscillation periods indicates retrograde propagation.
  The rms south-north travel-time shift $\delta\tau_{\rm rms}$ for each $m$ is obtained from the fitted Lorentzian $F_m(\omega)-B$ using Parseval's theorem with the window function taken into account.
  The rms horizontal flow speed of the r~modes along the equator in the near-surface layers, $v_{\rm rms}$, is converted from the $\delta\tau_{\rm rms}$ using conversion constants estimated from forward modeling (see text).
  A typical error in the $v_{\rm rms}$ is about 0.5~m~s$^{-1}$ based on the error propagation.
  The maximum flow speed along the equator for each mode is related to the rms flow speed by $v_{\rm max} = \sqrt{2}\,v_{\rm rms}$.
  Also listed are the mode frequencies from \citet{Loeptien2018} for comparison.
}
\end{table*}

\subsection{Estimates of r-mode velocity}
To obtain a rough estimate of the flow speed associated with a mode, first we compute in the ray approximation the travel-time shifts due to a prescribed toroidal flow field \citep[see, e.g.,][]{Saio1982}.
For the forward calculation, we choose a maximum horizontal flow speed at the surface of 2~m~s$^{-1}$.
We considered two toroidal flows: the first one is independent of depth and the second one decreases linearly with depth to vanish at 0.9~$R_\odot$.
The forward-modeled travel-time shifts in the north-south direction are Gaussian smoothed in longitude and latitude and averaged over travel distances in the same way as the measurements.
The resulting travel-time shifts decrease with increasing $m$ due to the smoothing in longitude.
The maximum travel-time shifts from the first flow model range from 0.13~s to 0.05~s depending on $m$, from which we derive conversion constants of 15.6--41.7~m~s$^{-2}$ to convert from travel-time shifts to the surface flow speed.
The conversion constants from the second model are in general larger by a factor of $\sim$1.6 for each mode.

Next, we obtain the rms travel-time shifts $\delta\tau_{\rm rms}$ from the fitted Lorentzian profile $F_m(\omega)-B$ for each mode using Parseval's theorem.
The effect of the incomplete data coverage and the spectral leakage is estimated by applying the same window functions and analysis to synthetic data, and is taken into account when computing the $\delta\tau_{\rm rms}$.

Last, we use the above conversion constants to convert from $\delta\tau_{\rm rms}$ to the surface rms flow speed $v_{\rm rms}$.
The conversion constants adopted are from the first model in which the flow is independent of depth, and thus the estimated $v_{\rm rms}$ are conservative.
The results for modes with $3 \leq m \leq 15$ are listed in Table~\ref{tab:fit}.
For these modes, the surface velocity is on the order of 1\,--\,3~m~s$^{-1}$, with larger values for $7 \leq m \leq 13$.
The modes with the lowest $m$ values have velocities of order 1\,--\,2~m$\,$s$^{-1}$, as assumed in the RV study by \citet{Lanza2019}.

\section{Discussion}

Using an independent helioseismic method and a different data set, we have confirmed the existence of the equatorial global Rossby waves reported by \citet{Loeptien2018}. We have extended the observations to deeper layers (down to $\sim$63~Mm) and to a total period of 21 years by combining SOHO and SDO observations.
The power spectra obtained from three seven-year periods covering cycles 23 and 24 all show signatures of r~modes for $3 \leq m=\ell \leq 15$.
The measured mode frequencies and linewidths are generally consistent with the granulation-tracking results of \citet{Loeptien2018}.
However, in our data, excess power is observed on the low-frequency side of the line profiles for $m \geq 10$ which leads to a systematic shift of the fitted mode frequencies with respect to \citet{Loeptien2018}.

In order to check if the spectral leakage comes into play, we applied the same window functions and Fourier analysis to synthetic data.
The resulting spectra do not show apparent frequency shifts, implying that the excess power is not due to the leakage of neighboring modes.
Because r-mode frequencies may vary over the solar cycle, we also examined separately the spectrum shown in Fig.~\ref{fig:pows}c computed over a similar period as that used by \citet{Loeptien2018}.
The above-mentioned excess power remains present, which means that this difference with \citet{Loeptien2018} is not due to the use of MDI data in earlier periods.

Systematic effects associated with surface magnetic field cause solar cycle variations in the travel-time measurements \citep{Liang2015a}.
We tried excluding the data points inside the active regions from the averaging of CCFs; however, the resulting power spectrum was dominated by noise as the masking procedure removed a considerable amount of pixels.
Even if we could resolve the systematic effects of the surface magnetic field, the local flows surrounding the active regions may still enter the travel-time measurements \citep[e.g.,][]{Gizon2001,Gizon2004b}.
We suspect that the excess low-frequency power at low $m$ is caused by active regions and associated local flows since the rotation rate of the active regions is rather close to zero in our chosen rotation frame.
Also the low-frequency power is stronger during the first (May 1996--April 2003) and third (May 2010--April 2017) periods when the solar activity is relatively higher.
However, we cannot strictly exclude the possibility that the low-frequency power results from large-scale convection.
Examination of the power spectra of travel-time measurements at different latitudes may help clarify this issue.

The errors in mode frequency measurements listed in Table~\ref{tab:fit} were estimated from Monte Carlo simulations.
These errors can also be estimated using Eq.~2 from \citet{Libbrecht1992} which relates the errors to the fitted parameters $A, B, \Gamma_m$ and the total observation time.
The estimation of the errors on mode frequencies from the two methods are consistent.
Our signal-to-noise ratio $A/B$ is smaller than the granulation-tracking observations of \citet{Loeptien2018}.
Twenty-one years of data analyzed using time-distance helioseismology give similar frequency error estimates as six years analyzed using granulation tracking.

It is interesting to compare the mode amplitudes measured here with those reported by \citet{Loeptien2018}.
To this end, we first computed the radial vorticity of a prescribed r-mode toroidal flow field as used in the forward modeling, from which we obtained the ratio of rms velocity to rms vorticity in the latitude range between $\pm20\degr$ for each mode.
We then used the ratio to convert the rms radial vorticity reported by \citet{Loeptien2018} to the rms velocity.
The rms velocity estimated from their rms vorticity is on the order of 1--2~m~s$^{-1}$ and is consistent with the $v_{\rm rms}$ in Table~\ref{tab:fit}.

\begin{figure}
  \resizebox{\hsize}{!}{\includegraphics{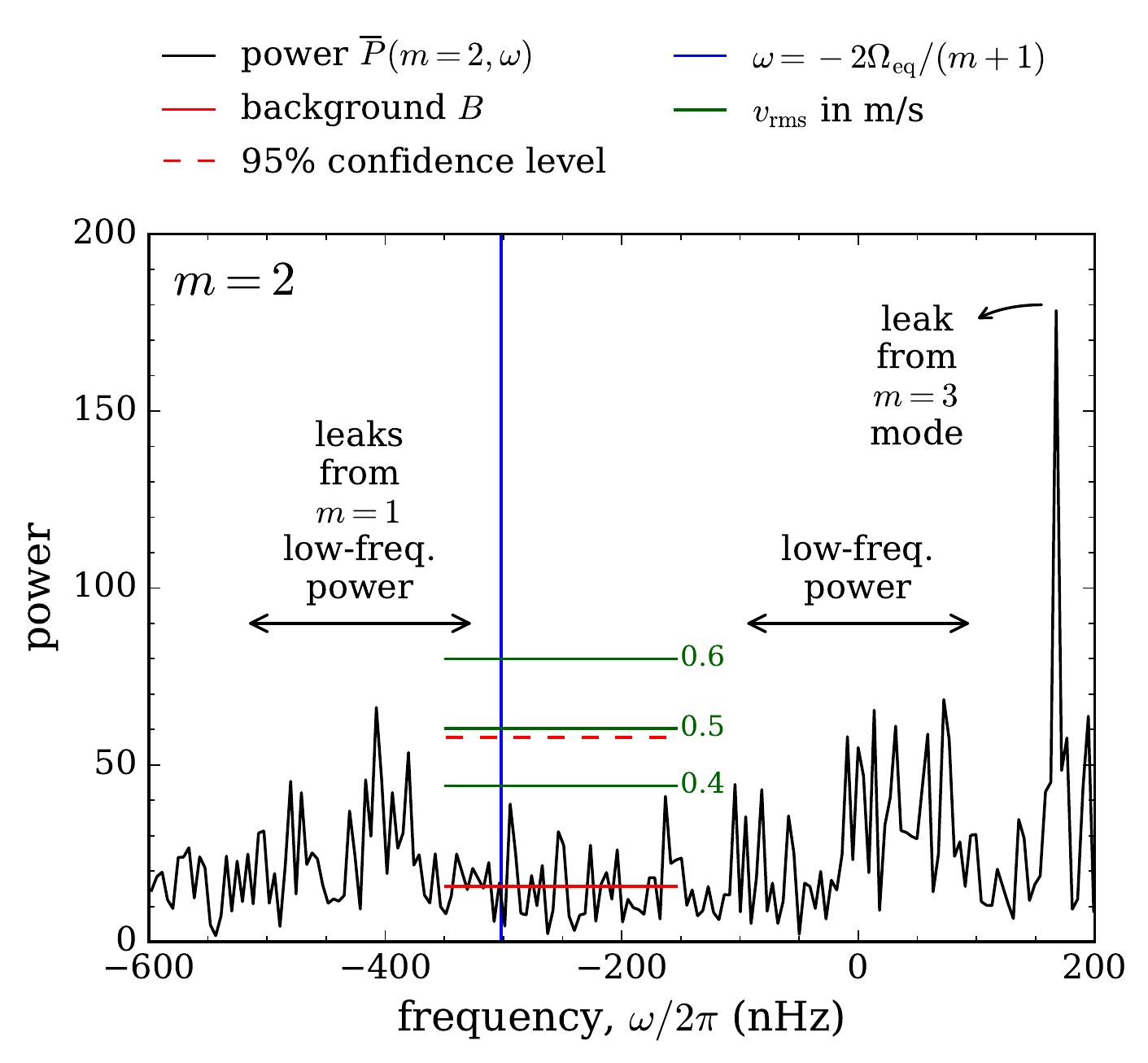}}
  \caption{ \label{fig:m2}
  Power spectrum of south-north travel-time shifts $\overline P(m, \omega)$ for $m=2$ (black solid line).
  The blue vertical line indicates the frequency of the classical $m=2$ sectoral r mode.
  The red solid line is the background $B$ estimated by a fit to the power in the frequency range between $-350$ and $-150$~nHz.
  The red dashed line is the threshold for 95\% confidence level; that is, the noise in the background only has a 5\% chance of being higher than this threshold for at least one frequency bin.
  We note that the spike around $-295$~nHz (on the right side of the blue line) is above the background but much lower than the 95\% confidence level. 
  The three green lines indicate the power that would correspond to a $m=2$ sectoral mode with rms velocity of 0.4, 0.5, or 0.6~m~s$^{-1}$.
  }
\end{figure}

We do not see any evidence for the $m=2$ mode (see Fig.~\ref{fig:m2}).
The travel-time measurements suffer from both time-independent and annual variations for $m \leq 2$ due to center-to-limb effects.
The $m=2$ mode is, however, expected to have a period of about 3/2 of the rotation period and should be cleanly separated from the center-to-limb systematics in the Fourier domain.
To place an upper limit on the velocity of a possible $m=2$ mode, we generated synthetic $m=2$ sinusoidal travel-time shifts with frequencies in the range between $-350$ and $-150$~nHz.
Three different amplitudes that correspond to rms velocity of 0.4, 0.5, and 0.6~m~s$^{-1}$ were implemented.
These synthetic data were added into the measurements and  Fourier analysis was applied with the window functions being taken into account.
Figure~\ref{fig:m2} shows the resulting power levels.
We find that an $m=2$ sectoral mode with $v_{\rm rms}< 0.5$~m~s$^{-1}$ would be difficult to identify in the power spectrum.

\begin{figure}
  \resizebox{\hsize}{!}{\includegraphics{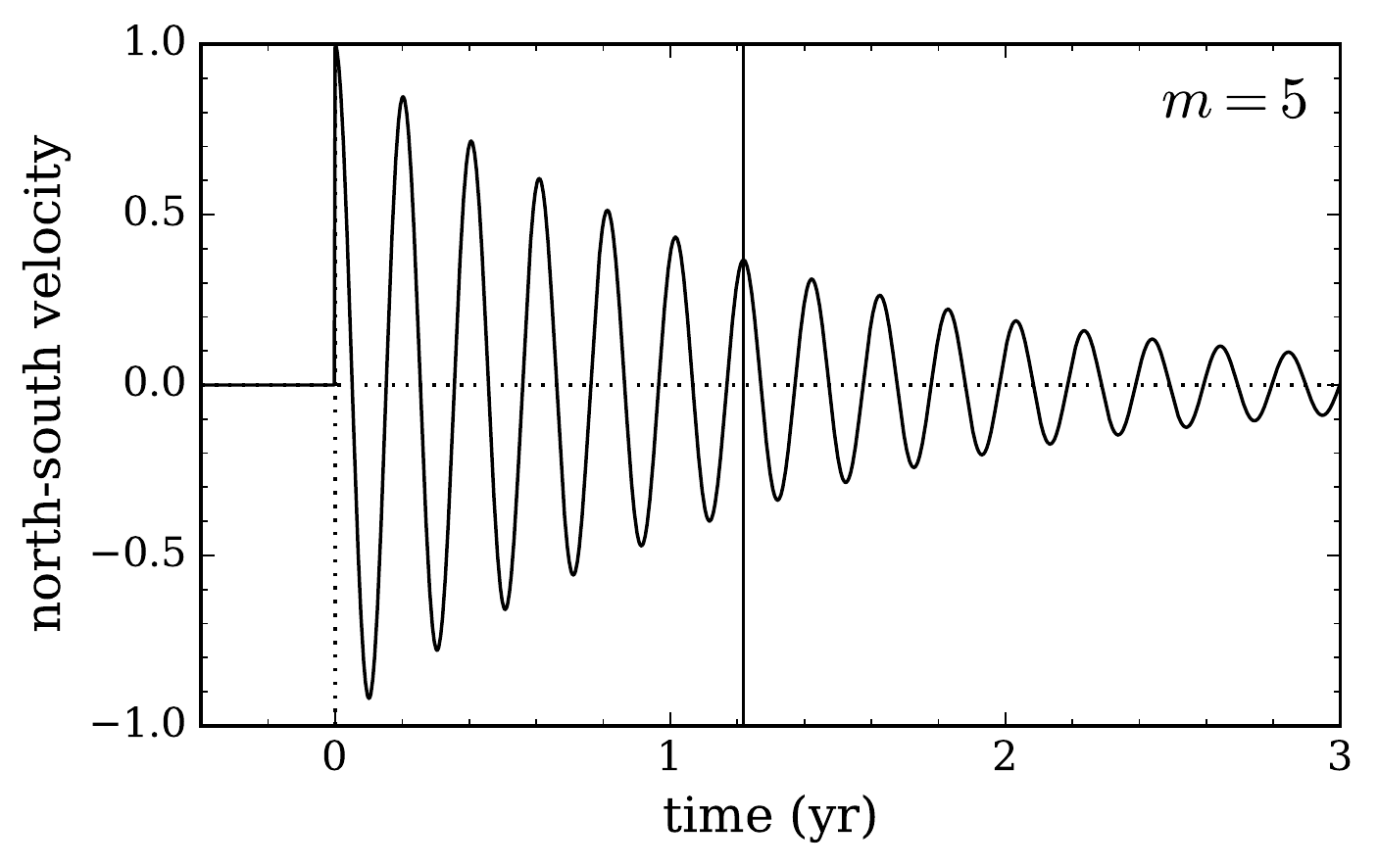}}
  \caption{ \label{fig:dt}
  Schematic damped oscillation of a sectoral r~mode with $m=5$, seen at a fixed longitude in the Sun's corotating frame.
  The value of the e-folding lifetime (vertical line) is as observed (see Table~\ref{tab:fit}).
  We note that if the oscillation were seen in the Earth's frame, the observed oscillation period would be about 6~days for $m=5$.
  }
\end{figure}

The e-folding lifetimes of r~modes, $2/\Gamma_m$, are on the order of a few months for most of the modes observed here.
The longest lived modes ($m = 3$ and $m = 5$) have lifetimes of more than one year.
Figure~\ref{fig:dt} shows a schematic excitation event (an exponentially decaying cosine) for the $m=5$ r~mode as seen in the corotating frame.
The physical mechanisms responsible for the excitation, or damping, of the r~modes are not yet known.

While the r-mode oscillations in the near-surface layers have been remarkably determined using the granulation-tracking and ring-diagram analysis, the depth dependence of solar r~modes throughout the convection zone is largely unknown.
Time-distance helioseismology \citep{Duvall1993} as used in this work has shown to be an important tool to study the r~modes deep inside the Sun.
More measurements of deep r-mode oscillations such as the one presented here will provide some insight into the nature of r~modes and, hopefully, further constrain theories of solar r~modes.

\begin{acknowledgements}
We thank B. L\"optien, B. Proxauf and J. Schou for useful discussions.
We also thank the referee for comments that helped improve the manuscript.
The HMI data used are courtesy of NASA/SDO and the HMI science team.
SOHO is a project of international cooperation between ESA and NASA.
The data were processed at the German Data Center for SDO (GDC-SDO), funded by the German Aerospace Center (DLR).
L.G. acknowledges partial support from the NYU Abu Dhabi Center for Space Science under grant G1502.
We used the workflow management system Pegasus funded by The National Science Foundation under OCI SI2-SSI program grant \#1148515 and the OCI SDCI program grant \#0722019.
\end{acknowledgements}

\bibliographystyle{aa}
\bibliography{ref}

\begin{appendix}
\section{Supplementary material}
\subsection{Inclination of the rotation axis} \label{app:dI}
Because the inclination angle of the Sun's rotation axis determined by \citet{Carrington1863} has been found to be slightly in error \citep{Beck2005,Hathaway2010}, corrections to both the instrument roll angle and the solar tilt angle $B_0$ are needed, otherwise a strong annual variation appears in the measured travel-time shifts and modulate the resulting spectrum.
Accordingly, an improvement to the mapping procedure used in \citet{Liang2018} is made by including corrections to the values of the two keywords \texttt{CRLT\_OBS} ($B_0$ angle) and \texttt{CROTA2} (instrument roll angle).
Following \citet[][, Eqs.3 and 4]{Larson2015}, the first order corrections are implemented as
\begin{align}
  \mathtt{CRLT\_OBS'} &= \mathtt{CRLT\_OBS} + \delta I \sin\left(2\pi \frac{t-t_\mathrm{ref}}{\rm 1~yr}\right), \\
  \mathtt{CROTA2'}\quad &= \mathtt{CROTA2} + \delta P + \delta I \cos\left(2\pi \frac{t-t_\mathrm{ref}}{\rm 1~yr}\right), \label{eq:dP}
\end{align}
where the primed keywords denote the updated values, $\delta I$ is the error in the inclination angle, $\delta P$ is an instrument-specific correction for CCD misalignment, $t$ and $t_\mathrm{ref}$ are the observation time and the time when \texttt{CRLT\_OBS} is close to zero, both expressed in years (365.25~days).
The $t_\mathrm{ref}$ is determined by fitting a sinusoid with a fixed period of 365.25~days to a time series of keyword \texttt{CRLT\_OBS}, and is 7~June 1996 01:19:34\_TAI for MDI data and 7~June 2010 14:17:20\_TAI for HMI data.
The adopted value of $\delta P$ for MDI data is 0.2$\degr$ \citep{Liang2017,Liang2018}, while that for HMI data is zero since the HMI team had calibrated and updated the value of \texttt{CROTA2} \citep{Couvidat2016,Hoeksema2018}.
As for $\delta I$, a value of $-0.08\degr$ from \citet{Hathaway2010} is adopted.
The effect of this correction is quantified by fitting a periodic function to the south-north travel-time shifts (see Appendix~\ref{app:bkg}).


\subsection{Weighting function} \label{app:wts}
\begin{figure}[h!]
  \resizebox{\hsize}{!}{\includegraphics{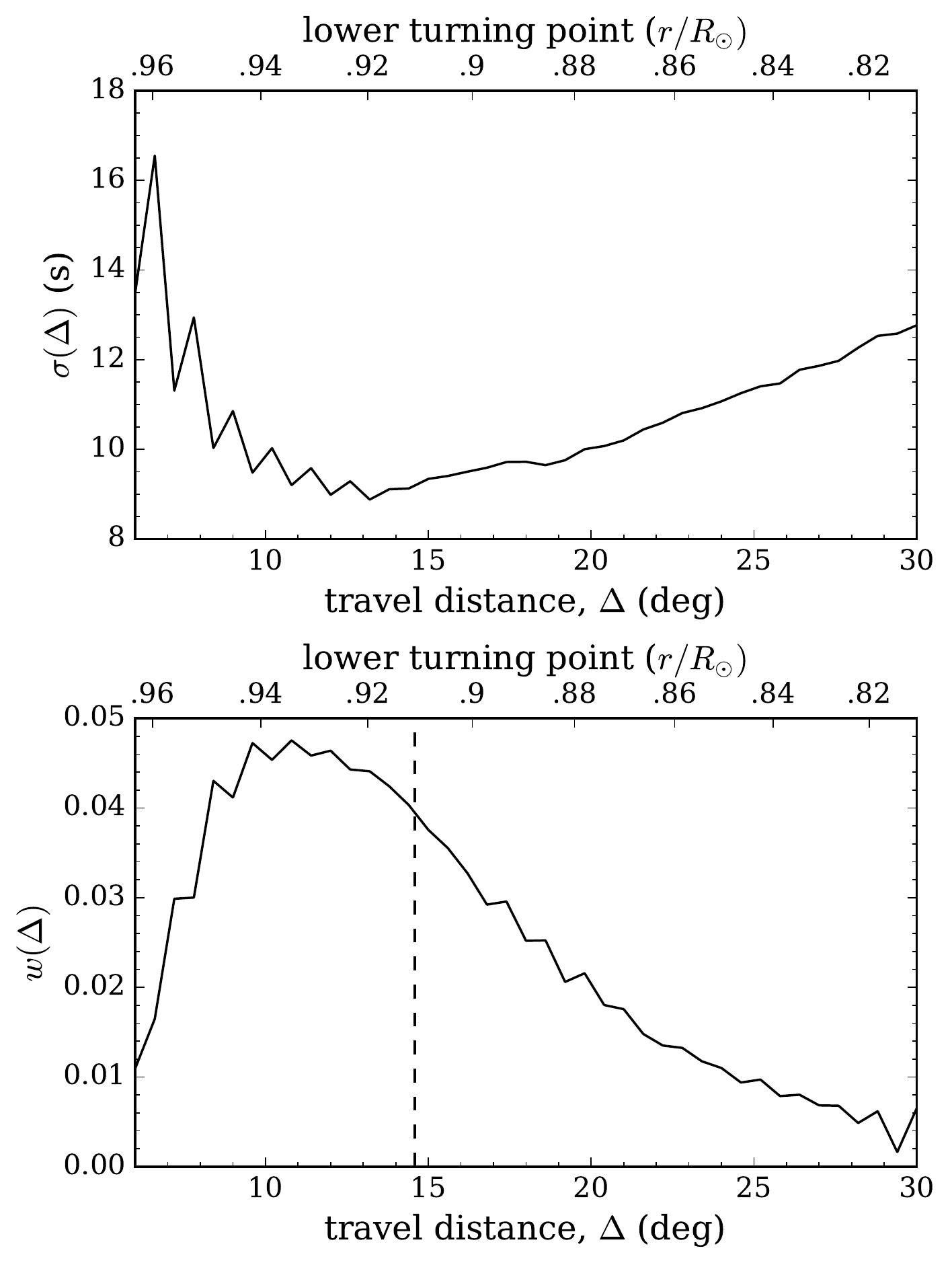}}
  \caption{ \label{fig:wts}
  \emph{Top}: Standard deviation of the measured travel-time shifts (square root of the diagonal elements of $\Lambda$) as a function of distance.
  The zigzag might be due to the discreteness of the windows that isolate the single-skip wavelet in the cross-covariance function when fitting the travel-time shifts, particularly for short-distance cases in which the slope of the single-skip ridge in the time-distance diagram is steepest.
  \emph{Bottom}: Weighting function, $w(\Delta)$, as a function of distance.
  The corresponding radii of the lower turning points from the ray approximation are indicated at the top.
  The weighted mean distance (vertical dashed line) is about 14.6$\degr$, which corresponds to a lower turning point of $\sim$0.91~$R_\odot$.
  }
\end{figure}

Here we use the notation introduced in Sect.~\ref{sec:td} and denote the measured south-north travel-time shifts by $\delta\tau^{(0)}(\psi, t, \Delta)$, where $\psi$ is the Carrington longitude, $t$ the observation time, and $\Delta$ the travel distance.
We have dropped the indices $i$ and $j$ in the subscripts of $\psi$ and $t$ for simplicity.
To combine the travel-time shifts for different travel distances, we follow the procedure in \citet[][, Sect.~5.1]{Jackiewicz2008}.
Given two random variables $X$ and $Y$, the covariance is defined by $\mathrm{Cov}[X, Y] = \langle XY \rangle - \langle X \rangle \langle Y \rangle$, where the angle brackets denote the expectation value (ensemble average).
The noise covariance matrix of the travel-time shifts between different distances is given by $\Lambda(\Delta, \Delta') = \mathrm{Cov}[\delta\tau^{(0)}(\psi, t, \Delta), \delta\tau^{(0)}(\psi, t, \Delta')]$.
The weighting for each distance is then given by
\begin{equation}
  w(\Delta) = \frac{\sum_{\Delta'}\Lambda^{-1}(\Delta, \Delta')}{\sum_{\Delta,\Delta'}\Lambda^{-1}(\Delta, \Delta')},
\end{equation}
and the weighted average of the measured travel-time shifts is
\begin{equation}
  \delta\tau^{(1)}(\psi,t) = \sum_{\Delta} w(\Delta) \, \delta\tau^{(0)}(\psi,t,\Delta).
\end{equation}

The square root of $\Lambda(\Delta,\Delta)$ is the standard deviation of the travel-time shifts, denoted by $\sigma(\Delta)$, which gives an estimate of the noise level at each distance.
Figure~\ref{fig:wts} shows the $\sigma(\Delta)$ and $w(\Delta)$ as a function of distance.
The trend of $\sigma$ is consistent with Fig.~2 of \citet{Beck2005} except that the noise level in their result is much lower probably owing to the use of phase-speed filter and different longitude and latitude ranges in the averaging.
For $\Delta < 10\degr$, the $\sigma$ increases rapidly because of the low spatial resolution of medium-$\ell$ Dopplergrams and thus lack of information about the acoustic waves residing in the near-surface layers.
Consequently, the weights for short distances are small.
For $\Delta > 10\degr$, the $\sigma$ increases with increasing distance because of geometrical spreading and damping.
Also, the noise correlation between different distances (off-diagonal elements of $\Lambda$) becomes greater when both $\Delta$ and $\Delta'$ are large \citep{Gizon2004}, which further reduces the weights for large-distance cases.


\subsection{Removal of the background variation} \label{app:bkg}

\begin{figure}
  \resizebox{\hsize}{!}{\includegraphics{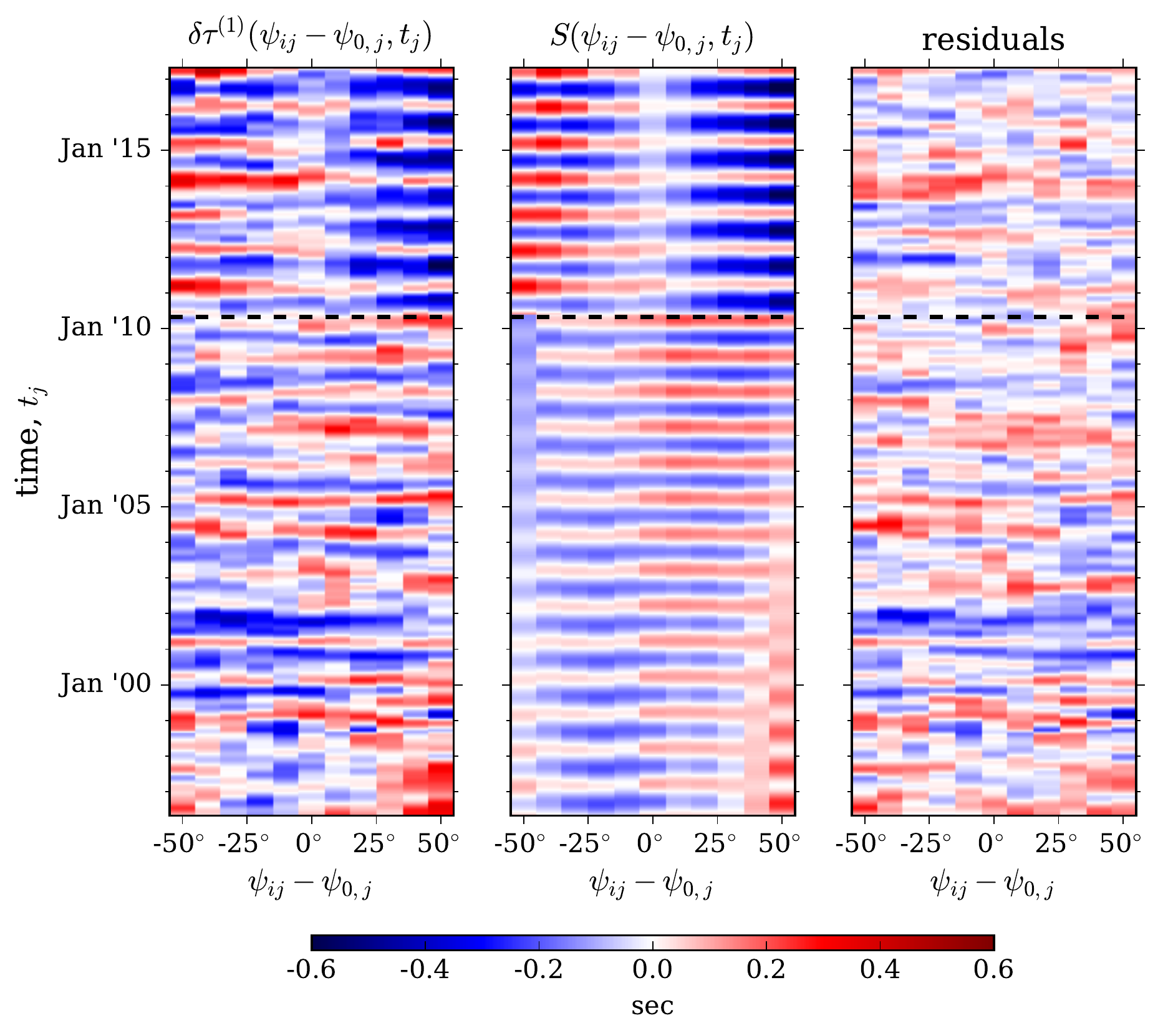}}
  \caption{ \label{fig:bkg1}
  \emph{Left}: travel-time measurements $\delta\tau^{(1)}(\psi'_{ij},t_j)$, Gaussian smoothed with $\mathrm{FWHM}=90$~days in time for better visualization, where $\psi'_{ij}\equiv\psi_{ij}-\psi_{0,j}$ is the separation from the central meridian.
  \emph{Middle}: fitted background $S(\psi'_{ij},t_j)$.
  \emph{Right}: difference between the two panels on the left.
  The horizontal dashed lines in all panels indicate the division between the results from MDI and HMI data; the MDI data used in this work span from May 1996 to April 2010 while HMI data span from May 2010 to April 2017.
  }
\end{figure}

\begin{figure}
  \resizebox{\hsize}{!}{\includegraphics{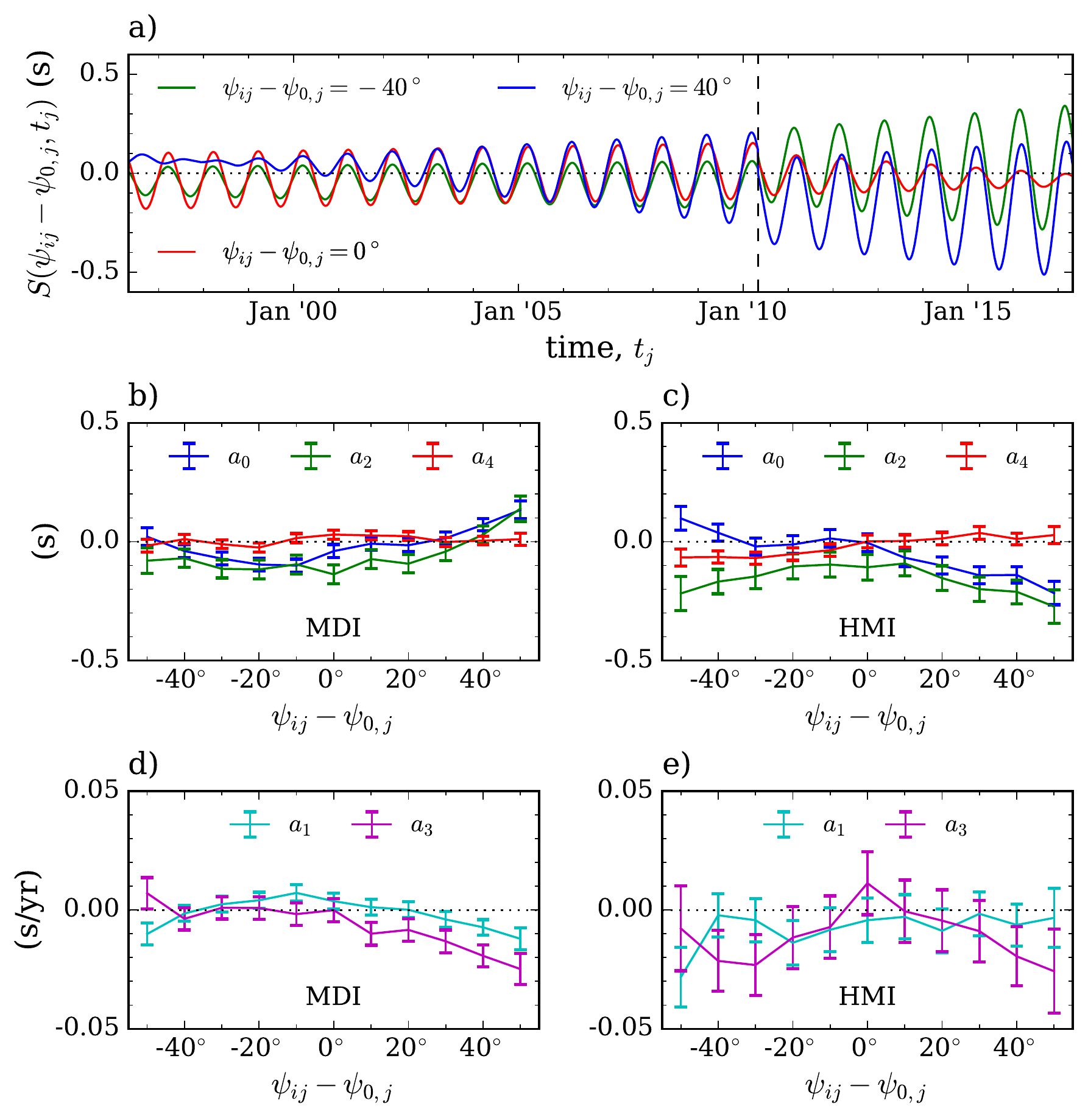}}
  \caption{ \label{fig:bkg2}
  Panel (a) shows the fitted $S(\psi'_{ij},t_j)$ at selected longitudes, where $\psi'_{ij}\equiv\psi_{ij}-\psi_{0,j}$ is the separation from the central meridian.
  The vertical dashed line indicates the division between the results from MDI and HMI data.
  Panels (b) to (e) show the fitted values of parameters as a function of longitude for the MDI and HMI data sets, respectively.
  The error bars are estimated from the covariance matrix of the fitting parameters.
  }
\end{figure}

Following the notation in Sect.~\ref{sec:td}, at time $t_j$, the Carrington longitudes of the travel-time measurements are denoted by $\psi_{ij}$, and the central meridian as seen by the observer is denoted by $\psi_{0,j}$.
Because the discussion here involves the center-to-limb effects, it is convenient to define the separation from the central meridian by $\psi'_{ij} \equiv \psi_{ij} - \psi_{0,j}$.
The weighted average of the travel-time shifts resulting from  Appendix~\ref{app:wts} is expressed in this coordinate system and denoted by $\delta\tau^{(1)}(\psi'_{ij}, t_j)$.

At each longitude $\psi'_{ij}$, a function representing the temporal variation of the background
\begin{equation} \label{eq:sin}
\begin{split}
  S(\psi'_{ij},t_j) &= a_0(\psi'_{ij}) + a_1(\psi'_{ij}) (t_j-t_{\rm ref}) \\
  &+ [a_2(\psi'_{ij}) + a_3(\psi'_{ij}) (t_j-t_{\rm ref})] \sin\left(2\pi \frac{t_j-t_{\rm ref}}{\rm 1~yr}\right) \\
  &+ a_4(\psi'_{ij})\cos\left(2\pi\frac{t_j-t_{\rm ref}}{\rm 1~yr}\right),
\end{split}
\end{equation}
is fitted to and subtracted from the $\delta\tau^{(1)}(\psi'_{ij},t_j)$ for MDI and HMI data separately.
Here $t_{\rm ref}$ is taken from Appendix~\ref{app:dI}.
The $a_0(\psi'_{ij})$ represents the time-independent background, and $a_1(\psi'_{ij})$ represents the linear component, if any, of the slowly varying background.
The $a_2(\psi'_{ij})$ accounts for an annual variation due to systematic errors such as the center-to-limb effects that vary with the $B_0$ angle (\texttt{CRLT\_OBS}); in addition, the $a_3(\psi'_{ij})$ is added since the magnitude of center-to-limb effects was found to be time dependent \citep{Liang2015b,Liang2018}.
The $a_4(\psi'_{ij})$ accounts for another annual variation caused by the error in the inclination of the rotation axis, if not completely removed in Sect.~\ref{app:dI}; the resulting error in the roll angle (i.e., Eq.~\ref{eq:dP}) may introduce a leakage of the solar rotation signal into the south-north travel-time shifts \citep{Beck2005}.

Figure~\ref{fig:bkg1} shows the $\delta\tau^{(1)}(\psi'_{ij},t_j)$, the fitted background $S(\psi'_{ij},t_j)$, and the residuals
\begin{equation}
    \delta\tau^{(2)}(\psi'_{ij},t_j) = \delta\tau^{(1)}(\psi'_{ij},t_j) - S(\psi'_{ij},t_j).
\end{equation}
The annual variation is clearly seen in the $\delta\tau^{(1)}(\psi'_{ij},t_j)$, suggesting the necessity of dealing with this systematic effect.
Also, the magnitude of $\delta\tau^{(1)}(\psi'_{ij},t_j)$ from HMI data at larger $|\psi'_{ij}|$ is systematically greater than the rest.
The fitted $S(\psi'_{ij},t_j)$ accounts for a large amount of the periodic variation of the background.
After removing the fitted background, the resulting $\delta\tau^{(2)}(\psi'_{ij},t_j)$ from the two data sets are more consistent with each other, and are ready for the Fourier analysis.

Figure~\ref{fig:bkg2} shows the $S(\psi'_{ij},t_j)$ at selected longitudes as well as the fitted values of the parameters as a function of longitude for the two data sets.
It is apparent that $a_0(\psi'_{ij})$ and $a_2(\psi'_{ij})$ (i.e., the center-to-limb effects) dominate the background variation while $a_4(\psi'_{ij})$ is nearly zero ($0.008\pm0.021$~s for MDI and $-0.015\pm0.038$~s for HMI within the longitude range $|\psi'_{ij}| \leq 30\degr$) since we have applied a correction $\delta I=-0.08\degr$ to the inclination angle of solar rotation axis as discussed in Appendix~\ref{app:dI}.
Without this correction, the value of $a_4(\psi'_{ij})$ is around 0.1--0.2~s.
We also tried $\delta I=-0.1\degr$ used by \citet{Larson2015} who rounded the value $-0.095\degr\pm0.002\degr$ determined by \citet{Beck2005} up to $-0.1\degr$; however, the $a_4(\psi'_{ij})$ becomes negative in this case ($-0.035\pm0.019$~s for MDI and $-0.058\pm0.038$~s for HMI).

The background variation in the measurements from HMI is larger than that from MDI since the magnitude of center-to-limb effects for HMI is greater \citep{Liang2017}.
The value of $a_2(\psi'_{ij})$ is expected to be negative because when the solar north pole is tilted toward the observer ($\mathtt{CRLT\_OBS} > 0$) the solar equator is on the southward side of the images where the center-to-limb variation in the south-north travel-time shifts is negative.

Furthermore, the background variation shows an east-west asymmetry which changes over time, especially in the measurements from MDI data.
The east-west asymmetry in the travel-time measurements from MDI was reported by \citet[][, Sect.~4.7.2]{Giles2000}.
He attributed the asymmetry to the nonuniform focus of the MDI camera, which changed at times during the mission.
We note that the granulation-tracking measurements also suffer from systematic errors similar to the center-to-limb effects, termed the shrinking-Sun effect \citep{Lisle2004}.
The shrinking-Sun effect not only depends on the distance to the disk center, but also exhibits an asymmetry in the east-west direction \citep{Loeptien2016,Loeptien2017}.
\citet{Loeptien2016} suggested that the east-west asymmetry in the shrinking-Sun effect could be caused by the influence of the solar rotation on the observing height.
If so, it would have a similar effect on the travel-time measurements since the magnitude of center-to-limb effects is strongly dependent upon the line-formation height of the observables \citep{Zhao2012,Baldner2012}.

\subsection{Power spectrum as seen from Earth} \label{app:earth}

\begin{figure}[h]
  \resizebox{\hsize}{!}{\includegraphics{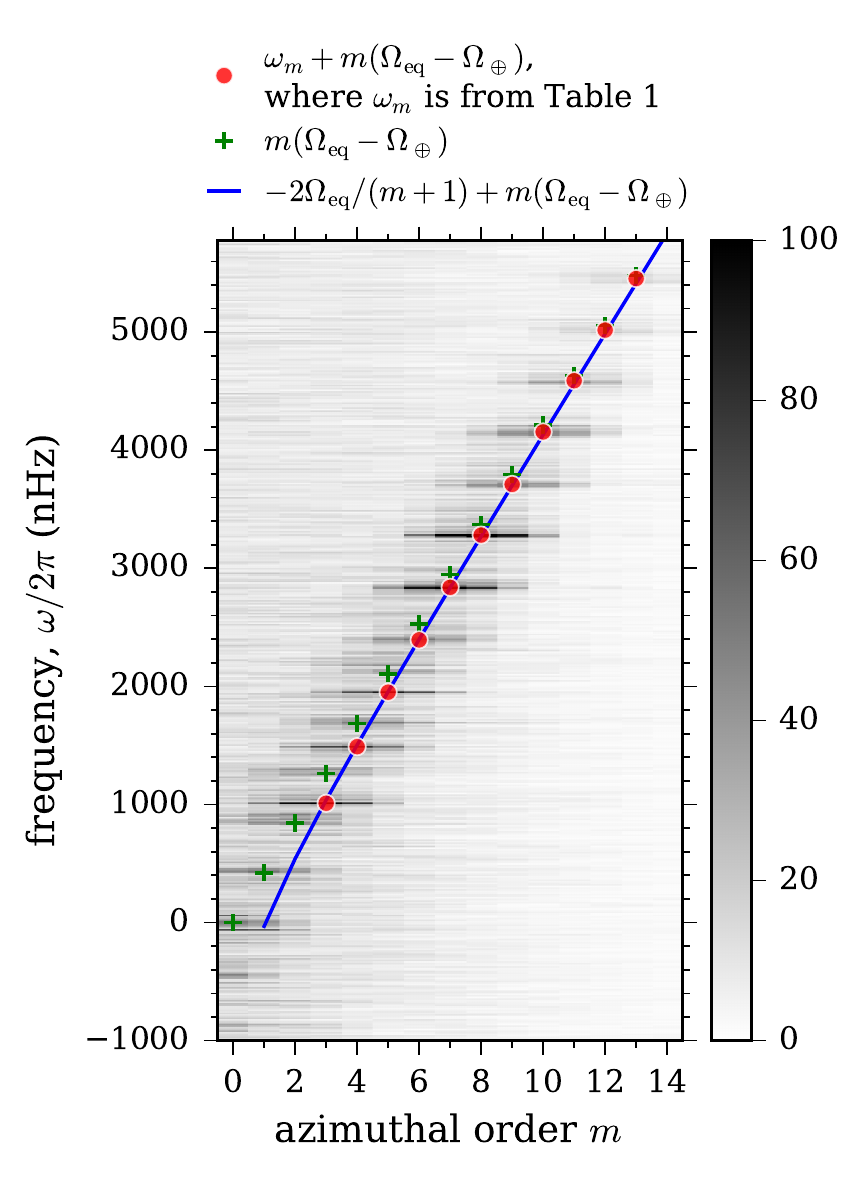}}
  \caption{ \label{fig:mw1}
  Power spectrum of south-north travel-time shifts computed in the Earth's frame.
  The blue line highlights the dispersion relation of sectoral modes of classical Rossby waves in this frame.
  The mode frequencies (red circles) are shifted by $m\times421.41$~nHz when measured in the Earth's frame.
  The low-frequency power from active regions or convection (near zero frequency at low $m$ in Figs.~\ref{fig:pows} and \ref{fig:zoom}) is also shifted to $m\times421.41$~nHz (green crosses) when measured in the Earth's frame.
  }
\end{figure}

Figure~\ref{fig:mw1} shows the power spectrum computed in the frame of the observer (Earth's frame).

\end{appendix}

\end{document}